# Dynamic control of defective gap mode through defect location

Lei Chang[1, 2], Yinghong Li[1], Yun Wu[1], Weimin Wang[1], Huimin Song[1]

[1]Science and Technology on Plasma Dynamics Laboratory, Xi'an 710038, China

[2]Southwest Institute of Physics, Chengdu 610225, China

**Abstract:**

A 1D model is developed for defective gap mode (DGM) with two types of boundary conditions: conducting mesh and conducting sleeve. For a periodically modulated system without defect, the normalized width of spectral gaps equals to the modulation factor, which is consistent with previous studies. For a periodic system with local defects introduced by the boundary conditions, it shows that the conducting-mesh-induced DGM is always well confined by spectral gaps while the conducting-sleeve-induced DGM is not. The defect location can be a useful tool to dynamically control the frequency and spatial periodicity of DGM inside spectral gaps. This controllability can be applied to optical microcavities and waveguides in photonic crystals and the interaction between gap eigenmodes and energetic particles in fusion plasmas.

**Introduction:**

It is well known that the propagation of electromagnetic waves in a periodic system is subject to Bragg's reflection and spectral gaps are formed within which this propagation is forbidden [1, 2]. Moreover, discrete gap modes can be formed with frequencies inside the spectral gaps if a local defect is introduced to break the system's periodicity and create an effective potential well to accommodate these modes [3-6]. The investigation of the defective gap mode (DGM) has been receiving great attention in multiple fields including photonic crystal [5, 7, 8] and fusion plasma [9-15]. For the photonic crystal, the DGM has promising applications in high-Q resonators [16-21], beam splitters [22, 23], sharp bend waveguides [24-28] and coupling structures [29-31], while for the fusion plasma, it plays an important role in confining energetic particles and thus generating fusion energy [32, 33]. However, little attention has been given to the effect of defect location on the DGM characteristics, and the types of boundary conditions which act as local defects were rarely focused on.

This paper first presents a 1D model developed for the DGM with two types of boundary conditions: conducting mesh and conducting sleeve. Then, it describes the effect of defect location on the DGM characteristics. It shows that the defect location can be used to dynamically control the DGM frequency and spatial periodicity inside spectral gaps. This may be of applications to optical microcavities and waveguides in photonic crystals and the interaction between gap eigenmodes and energetic particles in fusion plasmas.

**Analytical model:**

To study the propagation characteristics of electromagnetic waves in a periodically modulated system, we start from the Maxwell's equations:

$$\nabla \times \vec{E} = -\frac{\partial \vec{B}}{\partial t}, \quad \nabla \times \vec{B} = \mu \vec{j} + \mu\varepsilon \frac{\partial \vec{E}}{\partial t}. \tag{1}$$

Here, $\vec{E}$ and $\vec{B}$ are electric and magnetic wave fields respectively, $t$ is time, $\vec{j}$ is current density, $\mu$ is permeability, and $\varepsilon$ is permittivity. For a current-free ($|\vec{j}| = 0$) and charge-free ($\nabla \cdot \vec{E} = 0$) system considered here, the wave equation is therefore

$$\nabla^2 \vec{E} + \frac{\omega^2}{v_{ph}^2} \vec{E} = 0, \tag{2}$$

where $\omega$ is wave frequency and $v_{ph}$ is wave phase velocity. The relationship of $\mu\varepsilon = \mu_0 \varepsilon_0 \mu_r \varepsilon_r = \mu_0 \varepsilon_0 N^2 = 1/v_{ph}^2$ has been employed, with subscripts $0$ and $r$ denoting vacuum and relativity respectively and $N$ the index of refraction. Moreover, the wave form of $\exp(-i\omega t)$ is assumed with $i$ the imaginary unit. Here, we consider a 1D system with periodic modulation of the form

$$B_{0*} = B_0 [1 + \xi \cos(qz)] \tag{3}$$

with $B_{0*}$ modulated equilibrium, $B_0$ equilibrium, $\xi$ modulation factor, $2\pi/q$ modulation periodicity, and $z$ spatial coordinate. The modulated equilibrium introduces resonant coupling between the modes with $k = q/2$ and $k = -q/2$. This suggests the wave function of

$$E = A_+ \exp(iqz/2) + A_- \exp(-iqz/2) \tag{4}$$

where $A_+$ and $A_-$ are slow functions of $z$ compared to $\cos(qz)$, and $E$ is the scalar of $\vec{E}$. We assume $v_{ph} \propto B_{0*}$ (e. g. for Alfvénic waves) and make use of Taylor expansion ($\xi \ll 1$) to rewrite Eq. (2) for the constructed 1D periodic system

$$\frac{\partial^2 E}{\partial^2 z} + \frac{\omega^2}{v_{ph0}^2} [1 - 2\xi \cos(qz)] E = 0, \tag{5}$$

with $v_{ph0}$ the wave phase velocity for $B_0$. Through substituting Eq. (4) into Eq. (5) and separating harmonics of $\exp[i(k + q/2)]$ and $\exp[i(k - q/2)]$ (terms of the same harmonic should balance each other [14, 15]), we obtain two coupled equations:

$$\left[\frac{\omega^2}{v_{ph0}^2} - \left(k + \frac{q}{2}\right)^2\right] A_{+0} = \xi \frac{\omega^2}{v_{ph0}^2} A_{-0} \tag{6}$$

$$\left[\frac{\omega^2}{v_{ph0}^2} - \left(k - \frac{q}{2}\right)^2\right] A_{-0} = \xi \frac{\omega^2}{v_{ph0}^2} A_{+0} \tag{7}$$

where $A_+ = A_{+0} \exp(ikz)$ and $A_- = A_{-0} \exp(ikz)$ with $A_{+0}$ and $A_{-0}$ constants, and the relationship of $\cos(qz) = [\exp(iqz) + \exp(-iqz)]/2$ has been made use of. Terms of higher harmonics (e. g. $\exp[i(k + 3q/2)]$ and $\exp[i(k - 3q/2)]$), which are balanced by nonlinear parts of $E$, are neglected [14, 15]. Multiplying Eq. (6) with Eq. (7) results in the dispersion relation of electromagnetic waves propagating in a 1D periodic system,

$$\left[\frac{\omega^2}{v_{ph0}^2} - \left(k + \frac{q}{2}\right)^2\right]\left[\frac{\omega^2}{v_{ph0}^2} - \left(k - \frac{q}{2}\right)^2\right] = \left(\xi \frac{\omega^2}{v_{ph0}^2}\right)^2. \tag{8}$$

Equation (8) also tells the edges of the spectral gap between which waves are evanescent and thus cannot propagate longitudinally. By taking $k = 0$, we get

$$\omega_{eh} = \frac{q}{2} v_{ph0} (1 \mp \xi)^{-1/2} \text{ or } \omega_{el} = -\frac{q}{2} v_{ph0} (1 \mp \xi)^{-1/2} \tag{9}$$

with $\omega_{eh}$ the high-edge branch and $\omega_{el}$ the low-edge branch respectively, which further gives

$$\frac{\Delta\omega}{qv_{ph0}/2} = \frac{\Delta\omega}{\omega_0} \approx \xi. \tag{10}$$

Therefore, the normalized gap width is equal to the modulation factor, which is consistent with previous studies [14, 15, 34]. Here, $\omega_0$ is for $B_0$, standing for the center of the spectral gap.

To form a DGM inside the spectral gap, we consider two types of defects introduced by boundary conditions: conducting mesh ($E(z_0) = 0$) and conducting sleeve ($\partial E(z_0)/\partial z = 0$) with $z_0$ the defect location. The mesh and sleeve can be inserted across and wrapped around a plasma cylinder in experiments, respectively. These two boundary conditions will correlate $A_+$ and $A_-$ through Eq. (4) and create two types of defective gap modes. By substituting this correlation into Eqs. (6)-(7), we get the dispersion relation of conducting-mesh-induced DGM ($E(z_0) = 0$):

$$\frac{\omega^2}{v_{ph0}^2} - k^2 - \left(\frac{q}{2}\right)^2 = -\xi \frac{\omega^2}{v_{ph0}^2}\cos(qz_0), \tag{11}$$

$$kq = i\xi \frac{\omega^2}{v_{ph0}^2}\sin(qz_0), \tag{12}$$

and the dispersion relation of conducting-sleeve-induced DGM ($\partial E(z_0)/\partial z = 0$):

$$\frac{\omega^2}{v_{ph0}^2} - k^2 - \left(\frac{q}{2}\right)^2 = -\frac{\xi}{2}\frac{\omega^2}{v_{ph0}^2}\left[\frac{k+\frac{q}{2}}{k-\frac{q}{2}}\exp(iqz_0) + \frac{k-\frac{q}{2}}{k+\frac{q}{2}}\exp(-iqz_0)\right], \tag{13}$$

$$kq = \frac{\xi}{2}\frac{\omega^2}{v_{ph0}^2}\left[\frac{k+\frac{q}{2}}{k-\frac{q}{2}}\exp(iqz_0) - \frac{k-\frac{q}{2}}{k+\frac{q}{2}}\exp(-iqz_0)\right]. \tag{14}$$

Equations (11)-(14) describe the propagation characteristics of DGM for two types of boundary conditions, acting as local defects, and will be analyzed numerically in the following section.

Compared to the DGM models developed in [14, 15], which employs the wave equations of radially localized helicon mode and Alfvénic mode, respectively, together with a cold plasma dielectric tensor, the present model is based on Maxwell's equations only and is therefore more general. Moreover, it contains wave numbers so that making the study of the propagation characteristics of DGM possible. Additionally, it is 1D and does not consider radial structures, allowing easier experimental validations.

**Results and discussion:**
To illustrate the propagation characteristics of the conducting-mesh-induced DGM, we solve $\omega$ from Eq. (11) and Eq. (8) and plot the solutions together. Figure 1(a) shows the results. The employed conditions are typical parameters on the LAPD (LArge Plasma Device) [35], including equilibrium magnetic field strength of $B_0 = 0.12$ T, ion density of $n_i = 10^{18}$ m$^{-3}$, and ion specie of Helium. The equilibrium phase velocity of Alfvénic waves is as a result $v_{ph0} = B_0/\sqrt{\mu_0 n_i m_i} = 1.31 \times 10^6$ m/s with $m_i$ ion mass. Here, we choose $\xi = 0.5$ and $q = 40$ for clear illustration purpose. We observe that the dispersion curve of DGM bounces between the lowest and highest edges of the spectral gap as the defect location is varied, however, it is always well confined inside the spectral gap. Video 1(a) gives the details. This implies that the DGM frequency can be controlled accurately by adjusting the defect location carefully. Similarly, the propagation characteristics of the conducting-sleeve-induced DGM can be obtained by combining Eq. (13) and Eq. (8). However, as shown in Video 1(b), the dispersion curve of DGM is not always confined inside the spectral gap but depends on the defect location. There are resonance peaks around $|k| = 11.5$ and $|k| = 34.6$ for $z_0 = 0$ and $z_0 = 2\pi/2q$, respectively. Figure 1(b) shows the result for $z_0 = 2\pi/3q$. Therefore, the conducting-mesh-induced DGM has better confinement than the conducting-sleeve-induced DGM inside spectral gaps.

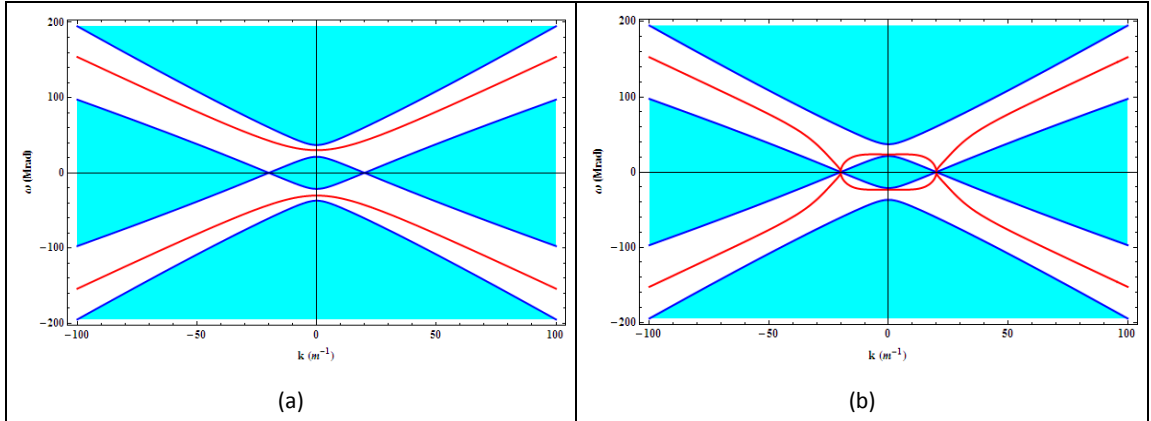

(a) (b)

Figure 1. Illustrations of conducting-mesh-induced DGM (a) and conducting-sleeve-induced DGM (b) inside spectral gaps for defect location $z_0 = 2\pi/3q$. DGM is labeled in red line while spectral gap in white area.

Next, to illustrate the propagation characteristics of DGM in a system with continuously changing periodicity (e. g. $q$ is increased from zero to infinity), we first set a constant wave number of $k = 20$ and keep other conditions the same as for Fig. 1. Figures 2(a) and 2(b) show the results of the conducting-mesh-induced DGM and the conducting-sleeve-induced DGM, respectively, for defect location $z_0 = 2\pi/3k$. Detailed variations of the DGM frequency with defect location can be found in Videos 2(a) and 2(b). We find that the defect location can be used to modulate the DGM spatial periodicity: modulation periodicity decreases with increasing $z_0$. Again, the conducting-mesh-induced DGM is always confined inside the spectral gap, whereas the conducting-sleeve-induced DGM is not. Then, we fix the wave frequency of $\omega = 1 \text{ Mrad/s}$ and keep other conditions the same as for Fig. 1. Figures 3(a) and 3(b) show the results of the conducting-mesh-induced DGM and the conducting-sleeve-induced DGM, respectively, for defect location $z_0 = 10$. For the conducting-mesh-induced DGM, the dependence of wave number on defect location is animated in Video 3. Again, we find that the spatial modulation periodicity decreases with $z_0$ increased. Moreover, the dispersion curve of the conducting-sleeve-induced DGM appears very noisy, indicating the possible existence of numerical uncertainties.

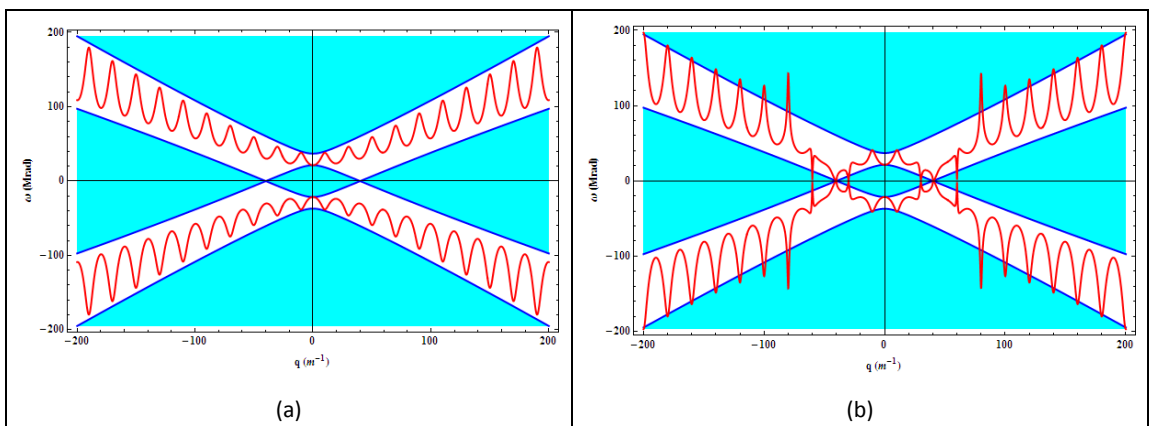

(a) (b)

Figure 2. Illustrations of conducting-mesh-induced DGM (a) and conducting-sleeve-induced DGM (b) inside spectral gaps for defect location $z_0 = 2\pi/3k$. DGM is labeled in red line while spectral gap in white area.

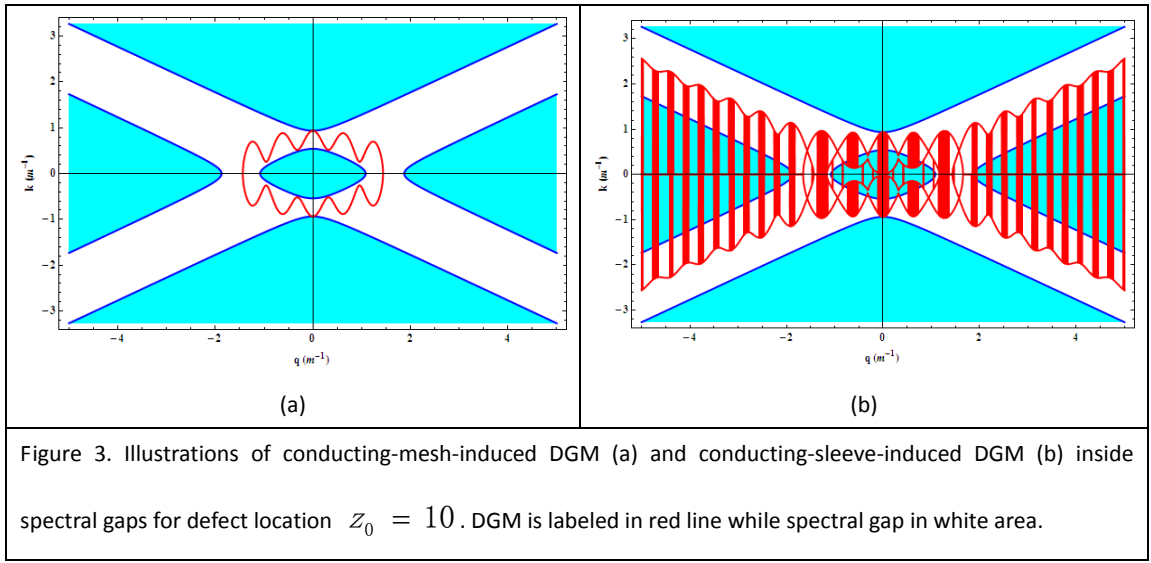

(a)        (b)

Figure 3. Illustrations of conducting-mesh-induced DGM (a) and conducting-sleeve-induced DGM (b) inside spectral gaps for defect location $z_0 = 10$. DGM is labeled in red line while spectral gap in white area.

The dynamic controllability of DGM frequency and spatial periodicity through defect location can be applied to optical microcavities and waveguides in photonic crystals. The idea is to adjust the DGM frequency and spatial periodicity dynamically to enhance the quality factor of resonant modes inside microcavities and to improve the impedance matching between defect and waveguides. It can also be employed to control the interaction between gap eigenmodes and energetic particles in fusion plasmas. During fusion reactions, weakly damped gap eigenmodes are readily destabilized by energetic ions and may degrade fast ion confinement [32, 33]. The idea is to break the wave-particle resonance condition to prevent energy transferability or to force the energy transferring direction, e. g. from gap eigenmodes to energetic particles, through varying the DGM frequency and spatial periodicity dynamically. For magnetized plasmas, the externally applied magnetic field powered by independent current coils can be a control tool to achieve dynamic modulation periodicity and defect location (e. g. increasing current on one of the coils).

**Conclusion:**

To show the propagation characteristics of DGM more explicitly, we developed a 1D analytical model based on Maxwell's equations only. Two types of boundary conditions are considered: conducting mesh and conducting sleeve. For a periodic system without defect, the normalized width of spectral gaps equals to the modulation factor, which is consistent with previous studies [14, 15, 34]. For a periodic system with local defects introduced by the boundary conditions, we find that the conducting-mesh-induced DGM has better confinement inside spectral gaps than the conducting-sleeve-induced DGM. Moreover, the defect location can be employed to dynamically control the frequency and spatial periodicity of DGM inside spectral gaps. This dynamic controllability has promising applications in the field of optical microcavities and waveguides in photonic crystals and the interaction between gap eigenmodes and energetic particles in fusion plasmas. The experimental validation of our model is in progress and will be addressed later.

**Acknowledgement:**

This project is supported by the National Natural Science Foundation of China through 11405271. Useful discussions with Boris Breizman, Matthew Hole and Yueqiang Liu are appreciated.